\documentclass[aps,prl, amsmath, showpacs, preprintnumbers,superscriptaddress, twocolumn,sort&compress,floatfix, amssymb]{revtex4}
\usepackage{graphicx}
\usepackage{rotating}
\usepackage{dcolumn}
\usepackage{bm}
\usepackage{color}
\usepackage{mathptmx, textcomp}
\usepackage[latin1]{inputenc}
\usepackage{braket}

\begin{document}

\author{M. J. Mark}
\affiliation{Institut f\"ur Experimentalphysik und Zentrum f\"ur Quantenphysik, Universit\"at Innsbruck, 6020 Innsbruck, Austria}
\author{E. Haller}
\affiliation{Institut f\"ur Experimentalphysik und Zentrum f\"ur Quantenphysik, Universit\"at Innsbruck, 6020 Innsbruck, Austria}
\affiliation{Institut f\"ur Quantenoptik und Quanteninformation, \"Osterreichische Akademie der Wissenschaften, 6020 Innsbruck, Austria}
\author{K. Lauber}
\author{J.~G. Danzl}\affiliation{Institut f\"ur Experimentalphysik und Zentrum f\"ur Quantenphysik, Universit\"at Innsbruck, 6020 Innsbruck, Austria}
\author{A. Janisch}
\author{H.~P. B\"uchler}
\affiliation{Institut f\"ur Theoretische Physik III, Universit\"at Stuttgart, Germany}
\author{A.~J. Daley}\affiliation{Department of Physics and Astronomy, University of Pittsburgh, Pittsburgh, PA 15260, USA}
\author{H.-C. N\"agerl}\affiliation{Institut f\"ur Experimentalphysik und Zentrum f\"ur Quantenphysik, Universit\"at Innsbruck, 6020 Innsbruck, Austria}

\title{Preparation and spectroscopy of a metastable Mott insulator state with attractive interactions}

\date{\today}

\pacs{37.10.Jk, 67.85.Hj, 03.75.Lm, 05.30.Rt}

\begin{abstract}
We prepare and study a metastable attractive Mott insulator state formed with bosonic atoms in a three-dimensional optical lattice. Starting from a Mott insulator with Cs atoms at weak repulsive interactions, we use a magnetic Feshbach resonance to tune the interactions to large attractive values and produce a metastable state pinned by attractive interactions with a lifetime on the order of 10 seconds. We probe the (de-)excitation spectrum via lattice modulation spectroscopy, measuring the interaction dependence of two- and three-body bound state energies. As a result of increased on-site three-body loss we observe resonance broadening and suppression of tunneling processes that produce three-body occupation.
\end{abstract}

\maketitle

Ultracold atomic gases in optical lattices provide a platform for investigating novel many-body dynamics in a highly controllable environment \cite{Bloch2008}. Recent studies of the quantum phase transition between a superfluid and an insulating Mott state for bosons \cite{Greiner2002,Gemelke2009,Fukuhara2009,Bakr2010,Sherson2010} and the metal to insulator transition for fermions \cite{Jordens2008,Schneider2008} exemplify the control available over these systems by tailoring the lattice potential, or tuning interparticle interactions using magnetic Feshbach resonances \cite{Chin2010}. This opens the door towards quantitative studies of phenomena that are not well understood in condensed matter physics, and also novel dynamics beyond what is normally realizable in solid state systems. Key examples of the latter include non-equilibrium dynamics such as transport processes \cite{Hackermuller2010, Trotzky2011}, quenches across phase transitions \cite{Cheneau2011,Chen2011,Kollath2007}, and the potential to realize metastable many-body states that are long-lived on experimental timescales. For example, repulsively bound atom pairs can be formed in experiments \cite{Winkler2006,Strohmaier2010} and are stable in optical lattice systems because of the lack of strong dissipative mechanisms such as lattice phonons that could remove energy on short timescales.

Here we prepare and study a novel metastable many-body state, specifically a metastable Mott insulator, in which particles are exponentially localized at individual sites through attractive two-body interactions. This state can be prepared via a sudden change in interparticle interactions, starting from a Mott insulator state with repulsive interactions, and then switching to attractive interactions abruptly on the tunneling timescale in the lattice. This is made possible by a broad Feshbach resonance for Cs atoms in the lowest hyperfine state \cite{Chin2004b}. We demonstrate that the resulting highly excited many-body state is long-lived, allowing for detailed studies of its properties. Using amplitude modulation (AM) of the optical lattice corresponding to two-photon Bragg transitions \cite{Stoeferle2004}, we measure the (de-)excitation spectrum. By identifying specific excitation resonances, we map out two-body bound state energies over a wide range of scattering lengths and make a quantitative comparison with the corresponding theoretical prediction \cite{Buechler2010}. We also observe shifted resonances connected to three-body bound states \cite{Johnson2009}, which feature a fast broadening and strength reduction for increasing attractive interactions due to increasing three-body loss \cite{Kraemer2006}. This reduction in strength corresponds to an inhibition of tunneling events that create three-body occupation due to on-site loss processes, which promises interesting effects on the many-body physics of the system \cite{Daley2009}.

The standard Bose-Hubbard (BH) model describes the dynamics of bosons in an optical lattice using the two-body on-site interaction energy $U$ and the nearest neighbor tunneling rate $J$. There are two groundstates for zero temperature, the superfluid state for $U\ll J$ and the Mott insulator for $U\gg J$. Here we prepare a metastable Mott insulator state with attractive interactions $(U<0)$. This state is analogous to a standard bosonic Mott insulator with repulsive interactions, in that particles are exponentially localized at different lattice sites, and the state exhibits an energy gap due to inter-particle interactions. However, for $U<0$ in a uniform system at unit filling, it is the most highly excited Hamiltonian eigenstate of the BH model in the limit $|U| \gg J$. The state is metastable on long timescales because of an energy gap of order $U$ to lower lying eigenstates, and there are no fast dissipative mechanisms to remove this energy from the system. This is analogous to metastability in a gas of repulsively bound pairs \cite{Winkler2006,Strohmaier2010} and the super-Tonks-Girardeau gas \cite{Haller2009}. The preparation of this highly excited system is realized by starting in a standard Mott insulator with $U>0$, $U\gg J$, and then rapidly switching $U$ to a large attractive value, on a timescale that is faster than $h/J$, but sufficiently slow to avoid population of higher Bloch bands.

We start with the production of an essentially pure Bose-Einstein condensate (BEC) of $1.0\times10^{5}$ Cs atoms in an optical dipole trap in the lowest hyperfine state following mainly the procedures detailed in Ref.\,\cite{Kraemer2004}. We load the BEC over the course of $500\,$ms into a cubic optical lattice created by three retro-reflected laser beams with a wavelength of $\lambda\,{=}\,1064.5\,$nm. The lattice depth $V_0$ for the measurements that follow is typically set to $V_0\,{=}\,20\,E_{\rm R}$, where $E_{\rm R}\,{=}\,h^2/(2m\lambda^2)$ is the atomic recoil energy with the mass $m$ of the Cs atom. The scattering length $a_{\rm S}$ can be tuned in a range between $\approx-2500\,$a$_{\rm 0}$ and $\approx1500\,$a$_{\rm 0}$ by applying a magnetic field $B$ between $0\,$G and $48\,$G, employing a broad Feshbach resonance with pole at $\sim-12\,$G and a narrow resonance at $\sim48\,$G \cite{Lange2009}. Here, a$_{\rm 0}$ is Bohr's radius. The zero crossing for $a_{\rm S}$ is at $17.119\,$G. During loading of the lattice the interaction strength is set to $a_{\rm S}\,{=}\,+220\,$a$_{\rm 0}$ to prepare an atomic Mott insulator state as the starting point for all subsequent measurements. The number of doubly occupied lattice sites can be controlled via the external confinement, which is primarily set by the dipole trap laser beams used for the initial BEC preparation.

\begin{figure}
\includegraphics[width=8.5cm]{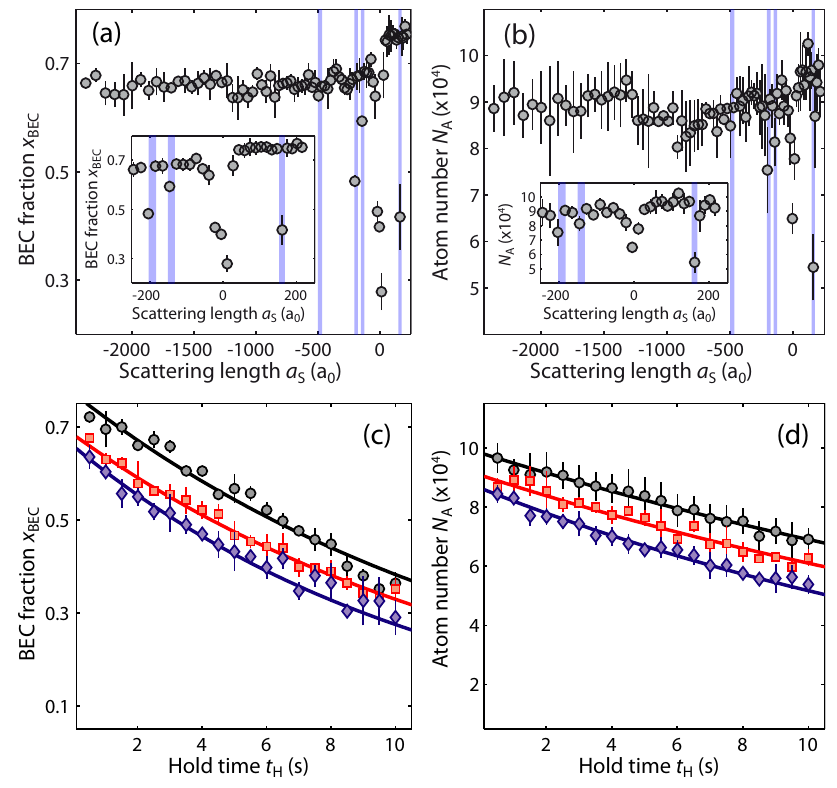}
\caption{(color online) Stability of the many-body system for attractive interactions a) BEC fraction $x_{\rm BEC}$ as a function of $a_{\rm S}$ with $t_{\rm H}\,{=}\,50\,$ms. The narrow shaded areas indicate the locations of Feshbach resonances. The inset shows the region around $a_{\rm S}\,{=}\,0\,$a$_{\rm 0}$. b) Number of remaining atoms $N_{\rm A}$ as a function of $a_{\rm S}$ with the same settings as in a). c) $x_{\rm BEC}$ as a function of $t_{\rm H}$ for $a_{\rm S}\,{=}\,+220\,$a$_{\rm 0}$ (circles), $a_{\rm S}\,{=}\,-240\,$a$_{\rm 0}$ (squares) and $a_{\rm S}\,{=}\,-2000\,$a$_{\rm 0}$ (diamonds). The solid lines are exponential fits giving $1/e$-decay times of $14.2\pm0.6\,$s ($+220\,$a$_{\rm 0}$), $13.7\pm0.6\,$s ($-240\,$a$_{\rm 0}$) and $11.4\pm0.4\,$s ($-2000\,$a$_{\rm 0}$). d) $N_{\rm A}$ as a function of $t_{\rm H}$ for the same settings as in c). The solid lines are exponential fits giving decay times of $28.3\pm0.9\,$s ($+220\,$a$_{\rm 0}$), $25.2\pm1.8\,$s ($-240\,$a$_{\rm 0}$) and $19.5\pm0.5\,$s ($-2000\,$a$_{\rm 0}$). The vertical error bars reflect the one-sigma statistical error. \label{Fig1}}
\end{figure}

To first investigate the stability of the many-body system with attractive interactions, we ramp $B$ and therefore the interaction strength to the desired value with a ramp speed of $2.5\,$G/ms, wait for a variable hold time $t_{\rm H}$ and return to the initial interaction strength. Subsequently we ramp down the optical lattice and recapture the cloud of particles in the dipole trap. Using the time-of-flight technique, we deduce the number of remaining atoms $N_{\rm A}$ and the BEC fraction $x_{\rm BEC}$. In the course of the ramps, we have to cross several narrow Feshbach resonances \cite{Chin2004b}. To avoid heating through interaction-induced band-transfer \cite{Koehl2005}, we cross the resonances using fast ramps with a speed of $2\times10^4\,$G/ms \cite{Danzl2009}. Figure\,\ref{Fig1}(a) and (b) show $x_{\rm BEC}$ and $N_{\rm A}$ as a function of $a_{\rm S}$ for $t_{\rm H}\,{=}\,50\,$ms. Sharp dips for both observables occur near the expected locations of the narrow resonances. Also, around the zero crossing of $a_{\rm S}$ the system becomes unstable. An approximately 20\% (10\%) decrease for $x_{\rm BEC}$ ($N_{\rm A}$) is observed for $a_{\rm S}\,{<}0$ compared to the values at $a_{\rm S}\,{>}0$, probably caused by the crossing of the zero interaction region within a finite time. Figure\,\ref{Fig1}(c) and (d) show $x_{\rm BEC}$ and $N_{\rm A}$ as a function of $t_{\rm H}$ for different values of $a_{\rm S}$. As one would expect, $x_{\rm BEC}$ decays somewhat faster than $N_{\rm A}$ at given $a_{\rm S}$. Exponential fits to the data yield lifetimes. These are reduced for more attractive interactions, but stay in the range of 10 to 20 seconds, more than $30$ times larger than the bare tunneling time $h/J$ at this lattice depth. This clearly shows that the system is stabilized by attractive interactions and allows us to investigate its properties in more detail.

\begin{figure}
\includegraphics[width=8.5cm]{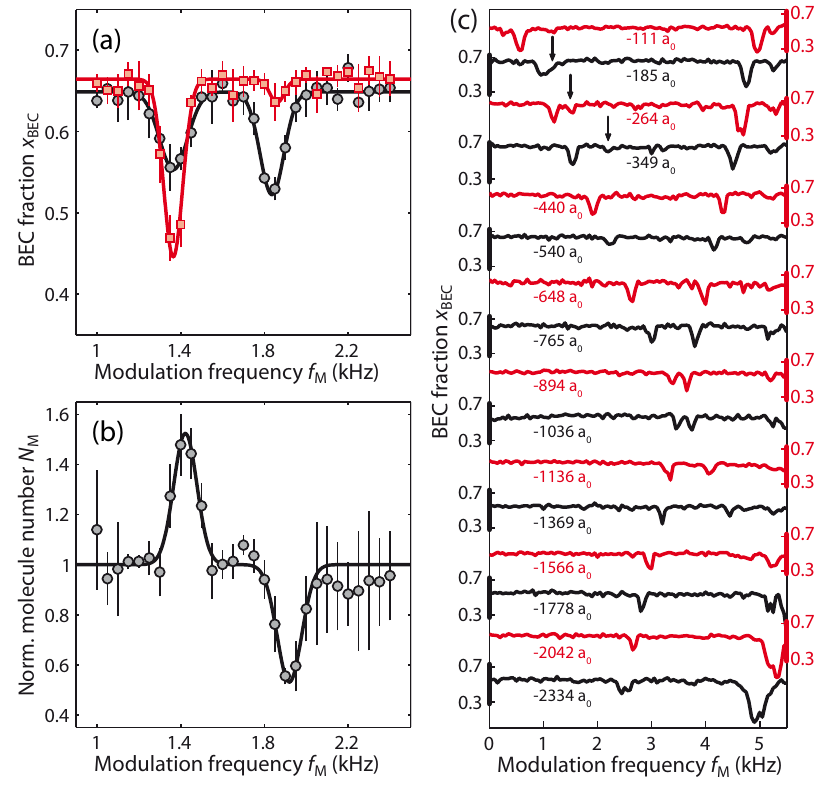}
\caption{(color online) Excitation spectrum for a Mott insulating state with attractive interactions a) BEC fraction $x_{\rm BEC}$ as a function of $f_{\rm M}$ for high (low) initial density shown as circles (squares) at $a_{\rm S}\,{=}\,-306\,$a$_{\rm 0}$. The solid lines are double Gaussian fits. b) Normalized molecule number $N_{\rm M}$ as a function of $f_{\rm M}$ at $a_{\rm S}\,{=}\,-306\,$a$_{\rm 0}$. The vertical error bars reflect the one-sigma statistical error. c) Set of excitation spectra with $x_{\rm BEC}$ as a function of $f_{\rm M}$ as $a_{\rm S}$ is varied: $a_{\rm S}\,{=}\,-111\,$a$_{\rm 0}$ (top) to $a_{\rm S}\,{=}\,-2334\,$a$_{\rm 0}$ (bottom). The three-body resonance is indicated by an arrow.\label{Fig2}}
\end{figure}

In this work we focus on the (de-)excitation spectrum \cite{Greiner2002,Stoeferle2004}, which we measure by AM at frequency $f_{\rm M}$ of one of the lattice beams at typically $20\,\%$ of $V_0$ for a duration of $t_{\rm H}\,{=}\,300\,$ms. Tunneling processes at $U<0$ to sites with non-zero occupation lower the overall energy and lead to a deexcitation of the system. The deexcitations - through the spatial configuration - are mapped onto excitations when returning subsequently to repulsive interactions, leading to an increased overall energy of the system, which we detect again by measuring $x_{\rm BEC}$ as a sensitive indicator for the energy deposited into the system. Figure\,\ref{Fig2}(a) shows the measured excitation spectrum for $a_{\rm S}\,{=}\,-306\,$a$_{\rm 0}$ in the vicinity of $|U|/h \approx 1.6\,$kHz as calculated from the standard BH model. A double resonance structure can clearly be identified, similar to our previous work in the regime of strong repulsive interactions \cite{Mark2011}. At comparatively low initial densities, with a small number of doubly occupied sites, the lower resonance is dominant, whereas for higher initial densities, giving a larger fraction of doubly occupied sites, the upper resonance becomes more pronounced at the expense of the lower resonance. As in the repulsive case, this splitting is caused by the energy difference for excitations in the different Mott shells through effective multibody interactions for three particles at the same lattice site \cite{Johnson2009}. We associate the lower resonance with excitations in the singly occupied shell, creating doubly occupied sites, i.~e. two-body bound states in the presence of the lattice, and the upper resonance with excitations in the doubly occupied shell, creating triply occupied sites, i.~e. three-body bound states in the presence of the lattice. In contrast to the repulsive case \cite{Mark2011} the three-body resonance is now at higher energies. Our interpretation is confirmed by a measurement of the number of doubly occupied sites through molecule formation \cite{Danzl2009}. Figure\,\ref{Fig2}(b) shows the molecule number $N_{\rm M}$ as a function of $f_{\rm M}$. For the lower resonance we find a greatly increased probability for dimer formation, which we attribute to an increase in the number of sites with double occupancy, while for the upper resonance we observe a strongly reduced probability for dimer formation, in agreement with the fact that doubly occupied sites are resonantly emptied upon modulation and that particles at triply occupied sites are lost due to fast three-body recombination at negative $a_{\rm S}$ \cite{Kraemer2006}.

\begin{figure}
\includegraphics[width=8.5cm]{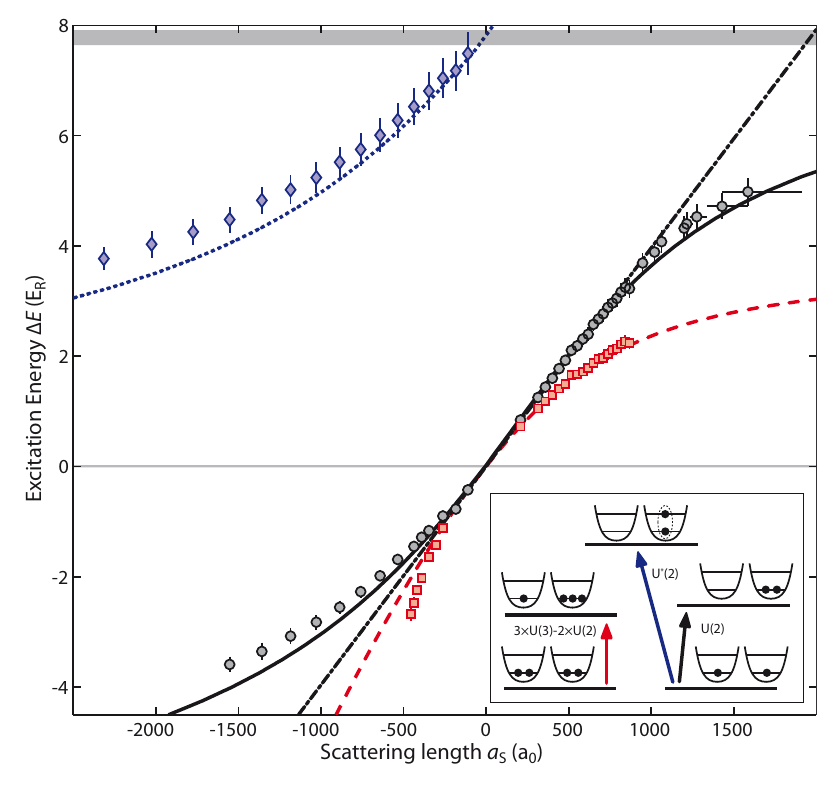}
\caption{(color online) Excitation energies $\Delta E$ into two- and three-body bound states. Two-body bound state excitation energy (circles), three-body bound state excitation energy (squares), and first excited two-body bound state excitation energy (diamonds, obtained by doubling the measured values) are shown as a function of $a_{\rm S}$. We have added to the plot the data for $a_{\rm S} > 0$ from Ref.~\cite{Mark2011}, augmented by new measurements for $a_{\rm S}>1000\,$a$_{\rm 0}$. The shaded areas at the top indicate the first two single-particle bands. The solid (dashed) line is the calculated excitation energies for the two-body (three-body) bound state and the dotted line is the calculated excitation energy for the first excited two-body bound state. The dashed-dotted line is the standard BH model calculation. The vertical error bars reflect the one-sigma statistical error as derived from the gaussian fit. The horizontal error bars indicate the variation of $a_{\rm S}$ over the cloud due to the gradient in $B$. The inset illustrates the main excitation processes.\label{Fig3}}
\end{figure}

Figure\,\ref{Fig2}(c) shows excitation spectra taken over a comparatively large frequency range as $a_{\rm S}$ is varied from $-112\,$a$_{\rm 0}$ to $-2334\,$a$_{\rm 0}$. The resonance discussed above corresponding to the excitation into the two-body bound state is clearly visible and it is shifted to higher frequencies as $|a_{\rm S}|$ is increased, as one would expect. The three-body resonance is only visible for weak interaction strengths. It also shifts to higher frequencies, as expected. We will discuss the behavior of this resonance below. Interestingly, we observe another resonance with an inverse behavior compared to the two-body and three-body resonances, with decreasing frequency as $|a_{\rm S}|$ increases. We observe an additional resonance at twice the frequency with the same behavior, visible in the spectra taken for strong interactions. These resonances do not disappear when we prepare a purely singly-occupied Mott insulator. We therefore suspect that they are related to excitations into the first excited two-body bound state of the lattice \cite{Buechler2010}. The strong visibility of the half-frequency resonance most likely is a result of the presence of the first harmonic in the frequency spectrum of $J$, which for our parameters reaches a strength of up to $20\,$\% of the main frequency component \cite{supmat}.

Figure\,\ref{Fig3} provides an overview over the measured excitation energies $\Delta E$ as $a_{\rm S}$ is varied and compares our data to the prediction by theory \cite{Buechler2010,Mark2011}. The three relevant excitation processes are illustrated in the inset to this figure. We determine $\Delta E = h f_{{\rm M},c}$ by fitting simple gaussians to the loss features as shown in Figure\,\ref{Fig2}(a) and taking the peak positions $f_{{\rm M},c}$. The data for repulsive interactions ($a_{\rm S}>0$) is taken from Ref.\,\cite{Mark2011}, augmented by new measurements at strong repulsive interactions ($a_{\rm S}>1000\,$a$_{\rm 0}$). In general, we find good agreement between our measurements and the calculated energies for the excitations to the lowest-band two-body bound state using the exact numerical results \cite{Buechler2010,supmat}, and the fitting function described in Ref.\,\cite{Mark2011}, which extrapolates the result for the three-body resonance from Ref.\,\cite{Buechler2010}. Only for comparatively strongly attractive interactions do we find a significant deviation for the energy of the three-body bound state, which is expected due to more complex three-body physics arising at negative values for the scattering length \cite{vStecher2011}. The excitation energies into the first excited two-body bound states, derived by doubling the measured frequency values, show the same qualitative behavior as the exact numerical calculations \cite{Buechler2010,supmat}, though with a significant offset for strong attractive interactions. Note that the exact numerical analysis is only valid for two particles or a very dilute system, and does not include the influence of a Mott insulating background. Its contribution will increase for stronger particle interactions and might account for the offset observed.
Also, we are not able to completely exclude the possibility of a systematic
deviation in the calculation of $a_{\rm S}$ in this magnetic field region.

\begin{figure}
\includegraphics[width=8.5cm]{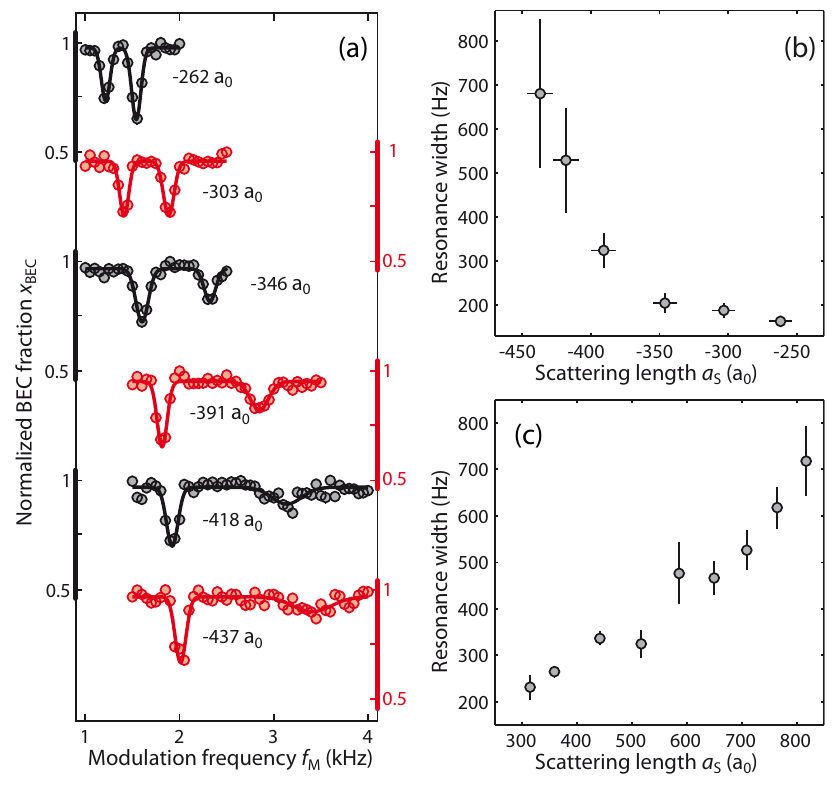}
\caption{(color online) Broadening of the three-body excitation resonance. a) Set of excitation spectra showing the normalized BEC fraction $x_{\rm BEC}$ as a function of $f_{\rm M}$ for various values of $a_{\rm S}$ as indicated. The resonance peak connected to the two-body (three-body) bound state is located at lower (higher) frequency. The solid lines are double-peaked Gaussian fits, from which the width (FWHM) of the resonances is deduced. b) FWHM of the three-body excitation resonance as a function of $a_{\rm S}$ for attractive interactions. c) For comparison, we show the FWHM of the three-body excitation resonance as a function of $a_{\rm S}$ for repulsive interactions \cite{Mark2011}.\label{Fig4}}
\end{figure}

Detailed spectra with the three-body bound state excitation resonance for attractive interactions are shown in Fig.\,\ref{Fig4}(a). Compared to the two-body bound-state resonance at lower frequencies, we clearly observe a broadening of the three-body resonance as $a_{\rm S}$ is increased. In fact, the resonance broadens so much that it finally disappears. Given our finite signal-to-noise ratio, we are not able to detect the resonance for $|a_{\rm S}|>500\,$a$_{\rm 0}$. Fig.\,\ref{Fig4}(b) shows the width (FWHM) as a function of $a_{\rm S}$. We note that, as there is currently no theory on the lineshapes of three-body resonances, the FWHM is obtained from simple gaussian fits as used above. The broadening leads to a decrease in the height of the resonance, i.e., a decrease in the formation rate of triply-occupied sites, and thus to a decrease in the overall loss rate. We interpret this behavior in terms of the process discussed in Ref.\,\cite{Daley2009}: Fast three-body loss can suppress the formation of triply occupied sites, analogous to the suppression of inelastic two-body processes as reported in Ref.\,\cite{Syassen2008} for the case of  weakly-bound molecules. As the three-body recombination rate scales like $\dot{N}\sim a_{\rm S}^4n^3$ with the density $n$, one would expect a strong increase of the three-body loss rate. We note that for repulsive interactions a similar broadening of the three-body bound-state excitation resonance can be observed. This is shown in Figure\,\ref{Fig4}(c). In this case the increase of the FWHM as a function of $a_{\rm S}$ is far less drastic. Further experimental and theoretical investigations will be needed for a quantitative analysis of the broadening, including a calculation of the modified on-site density of three-body bound states and the corresponding tunneling rates at attractive and repulsive interactions and the role of Efimov physics in confined dimensions \cite{Portegies2011}.

In summary we have investigated a metastable Mott insulating state with attractive interactions and have shown that this state exhibits a very long lifetime on the order of 10 seconds. By measuring excitation spectra we were able to determine the energies of two-body and three-body bound states. The broadening of the excitation resonance corresponding to the three-body bound state gives a strong indication that very high three-body recombination rates suppress the creation of triply occupied sites, thereby reducing the effective loss rate of the system, which also can be interpreted in terms of the quantum Zeno effect \cite{Misra1977} with the three-body recombination rate playing the role of the measurement. This effect can be used to realize a Bose-Hubbard model with a three-body hard-core constraint \cite{Daley2009}, leading to interesting many-body physics, including a dimer superfluid phase and a continuous supersolid phase for attractive bosons \cite{Diehl2010,Bonnes2011}, the realization of Pfaffian-like states in one-dimensional geometries \cite{Paredes2007}, and the stabilization of an atomic color superfluid for fermions \cite{Kantian2009}. The Mott insulating state with attractive interactions can also serve as starting point for the preparation of stable superfluid condensates at finite momentum with negative temperature \cite{Rapp2010}.

We are indebted to R. Grimm for generous support. We thank J. von Stecher, P. Johnson, and E. Tiesinga for fruitful discussions. We gratefully acknowledge funding by the Austrian Science Fund (FWF) within project I153-N16 and within the framework of the European Science Foundation (ESF) EuroQUASAR collective research project QuDeGPM.

\bibliographystyle{apsrev}

\clearpage
\newpage

\section{Supplementary material:
Preparation and spectroscopy of a metastable Mott insulator state with attractive interactions}

\subsection{Modifications and frequency components of $U$ and $J$ during amplitude modulation}

\begin{figure}[hbt]
\includegraphics[width=8.5cm]{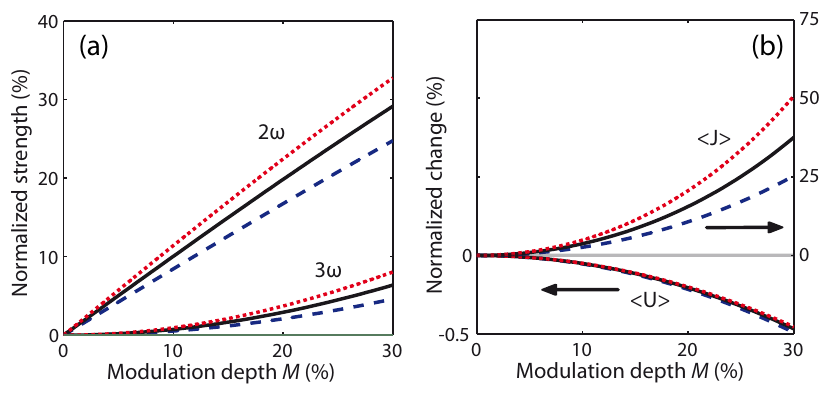}
\caption{(color online) a) Normalized strength of the $2\omega$ and $3\omega$ frequency components of $J(t)$ as a function of the modulation depth $M$ for $V_0=20\,$E$_{\rm R}$ (solid line), $V_0=15\,$E$_{\rm R}$ (dashed line), and $V_0=25\,$E$_{\rm R}$ (dotted line). b) Normalized change for the time averaged values $\langle U \rangle$ and $\langle J \rangle$ for $U$ and $J$ as a function of $M$ for different values of $V_0$ as in a). \label{Fig5}}
\end{figure}

A sinusoidal modulation of the lattice depth leads to a modulation in both $U$ and $J$. Excitations are driven mainly by the modulation of $J$ \cite{Clark2006s}. To characterize this process in a first approximation, we write the lattice depth modulation as $V(t)=V_0(1+M\sin(\omega_{\rm M} \, t))$ with $\omega_{\rm M}=2 \pi f_{\rm M}$ and numerically calculate $U(t)$ and $J(t)$ for each time step using tunnelling rates and Wannier functions for the lowest Bloch band. Here, we do not include the modification of $U$ and $J$ arising for stronger interactions. To derive the frequency components of $J(t)$, we Fourier transform $J(t)$, omit the DC part and normalize the strength of the frequency components to the peak strength at the main frequency $\omega_{\rm M}$. Due to the nonlinear dependence of $J$ on $V_0$, higher harmonics at $2\omega_{\rm M}$ and $3\omega_{\rm M}$ are visible, and their strength as a function of the modulation depth $M$ is shown in figure\,\ref{Fig5}(a) for three different values of the lattice depth $V_0$. For our experimental parameters ($V_0=20\,$E$_{\rm R}$, $M=20\,$\%) the $2\omega_{\rm M}$ component reaches $20\,$\% peak strength of the main frequency peak, whereas the $3\omega_{\rm M}$ component stays below $3\,$\%. This explains the good visibility of the half-frequency resonance into the first excited two-body bound state. Note that a half-frequency resonance is also present for the lowest two-body bound state as shown in Ref.\,\cite{Mark2011s}. The weak $3\omega_{\rm M}$ component would give rise to a further resonance, which, however, is not observable given the finite signal-to-noise ratio for our experiment.

\begin{figure}[hbt]
\includegraphics[width=8.5cm]{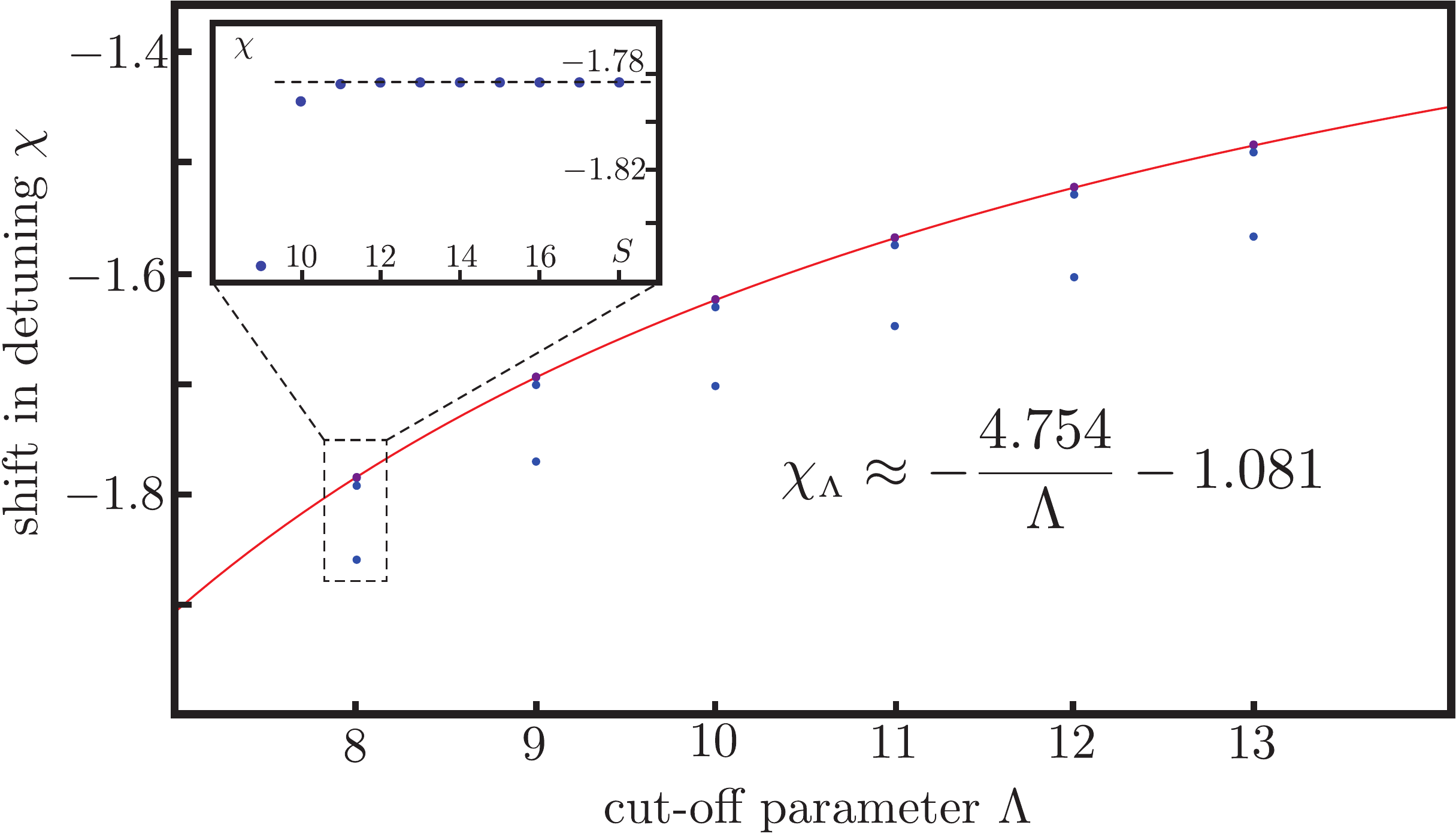}
\caption{(color online)  Convergence of the numerical analysis for increasing cut-off parameter and number of Bloch bands: Regularized shift of the detuning $\chi_{\Lambda}(E)$ at energy $E= 5.5 E_{\rm R}$ and for an optical lattice with $V_0=16 E_{\rm R}$. The dots are the values for different shell parameters $S$. The shift satisfies the asymptotic behavior $\chi_{\lambda} = c/\Lambda +\chi$ (solid line). The inset shows the fast convergence for fixed $\Lambda$ by increasing the shell parameter $S$, i.e., the involved number of Bloch bands.  \label{Fig6}}
\end{figure}

Due to the nonlinearity of $J$ and $U$ as a function of $V_0$ the corresponding time averaged values during the modulation are different from the values that one obtains assuming no modulation. Figure\,\ref{Fig5}(b) shows the relative change of the time averaged $\langle U \rangle$ and $\langle J \rangle$ as a function of $M$ for three different values of $V_0$. The shift of $U$ by less than $0.5\,$\% is nearly independent of $V_0$ and can be neglected in view of the uncertainties in our experiment. The shift of $J$ is more pronounced, but it does not affect the experimental results on the measurement of interaction energies. Nonetheless, this effect has to be taken into account for a future investigation of modulation-assisted tunneling rates, additionally to the changes due to modifications in the on-site wave functions arising from interactions \cite{Luehmann2011s}.

\subsection{Numerical analysis of the two-particle bound states}

The numerically efficient and precise determination of the attractive and repulsive bound state energies for two particles in a three-dimensional optical lattice has been presented in Ref.~\cite{Buechler2010s}. Here, we use this method for the comparison with the experimental data; the notation is in analogy to Ref.~\cite{Buechler2010s}. Within the two channel description of the short range pseudo-potential, the s-wave scattering length $a_{\rm S}$ is related to the detuning $\nu$ of the molecular channel, and the coupling $g$ via $4 \pi \hbar^2 a_{\rm S}/m = -  g^2/\nu$. The influence of the optical lattice is a shift $\chi(E)$ in the detuning and is associated with a change of the free particle properties due to the formation of Bloch bands. The relation between the bound state energies $E_{\rm \scriptscriptstyle bs}$ and the shift $\chi(E)$ in the detuning takes the form  $\chi(E_{\rm \scriptscriptstyle bs})=- \pi a/(8 E_{\rm R} a_{\rm S})$ with $a$ the lattice spacing. The numerical determination of $\chi(E)$ involves the precise determination of Bloch wave functions and band energies, as well as a summation over a high number of Bloch bands. In addition, it  requires a regularization of the coupling between the open and closed channel. Here, we choose the regularization $\alpha({\bf r}) = \int_{v(\Lambda)} d{\bf k} \exp(i {\bf k} {\bf r})/(2 \pi)^3$, where the volume $v(\Lambda) =  \Lambda^3 v_{0}$ is centered around ${\bf k}=0$ with $v_{0}= (2 \pi)^3/a^3$ the volume of the first Brillouin zone. For fixed cut-off $\Lambda$, the summation over Bloch bands converges very quickly for $S> \Lambda$, see inset to Fig.~\ref{Fig6}; the shell parameter  $S$ denotes the number of Bloch bands for each spatial direction included in the summation. Finally, we can remove the cut-off $\Lambda$ via the asymptotic scaling relation $\chi_{\Lambda}(E) =  c/\Lambda + \chi(E)$ \cite{Note1}, see Fig.~\ref{Fig6}. The convergence of the numerical data is checked by varying the shell parameter $S$, cut-off $\Lambda$, and the precision for the determination of the Bloch wave functions and Bloch bands.

Using the above procedure, the repulsive and attractive bound states are determined for the experimentally relevant parameters with $V_0=20 E_{\rm R}$. The lowest bound state for weak interactions is dominated by the contributions of two particles in the lowest Bloch band, and for increasing interactions smoothly connects to the state with a single bound molecule of mass $2m$ within the lowest Bloch band of an optical lattice with strength $2 V_0$. In turn, the first excited bound state corresponds for weak interactions
to a state with one particle in the lowest Bloch band and the second particle in the first excited Bloch band. Again for increasing interactions, this state smoothly connects to a single molecule in the first excited Bloch band.

\end{document}